\documentclass[twocolumn]{aastex631}

\usepackage{mathtools}
\usepackage{comment}
\begin{document}

\title{{Pre-Supernova Multiple Giant Eruptions in Massive Stars}}

\shorttitle{Multiple Eruptions in Massive stars}
\shortauthors{B. Mukhija and A. Kashi}

\author[0009-0007-1450-6490]{Bhawna Mukhija}
\affiliation{Department of Physics, Ariel University, Ariel, 4070000, Israel}
\email{bhawnam@ariel.ac.il}

\author[0000-0002-7840-0181]{Amit Kashi}
\email{kashi@ariel.ac.il}
\affiliation{Department of Physics, Ariel University, Ariel, 4070000, Israel}
\affiliation{Astrophysics, Geophysics, and Space Science (AGASS) Center, Ariel University, Ariel, 4070000, Israel}

\begin{abstract}

\end{abstract}


\begin{abstract}

Massive stars can exhibit giant eruptions with high mass loss shortly before their explosion as a core-collapse Supernova. These multiple giant eruptions (MGEs) may have a commutative effect that brings the star to a different state, possible one that favors the explosion. To address this problem we evolve a $\rm 100~M_{\odot}$ star and initiate a series of three giant eruptions lasting one year each, testing different mass loss rates and different metallicities. Following each eruption, we track the recovery phase to examine the post-eruption behavior of the star and its recovery timescale. The MGEs leads to a decrease in luminosity, accompanied by a slight increase in temperature. Later, during the recovery phases as the star starts to retain its equilibrium state, its luminosity increases.
The recovery time-scale varies significantly after each eruption for independent on the mass loss rate, but it is shorter for lower metallicities. For the higher mass-loss rates during the recovery phase, the outer layers of the star exhibit oscillations and undergo compression at higher metallicity. These oscillations are most likely a consequence of thermal imbalance in the outer envelope. This behavior at higher mass-loss rates also suggests that the thermal readjustments during recovery may create favorable conditions for a subsequent eruption of the star.

\end{abstract}

\section{Introduction} \label{sec:intro}

The Luminous Blue Variable (LBV) phase, characterized by episodic mass-loss events, is believed to play a crucial role in the overall mass-loss evolution of massive stars  \citep[e.g.,][]{1994PASP..106.1025H,1997ARA&A..35....1D, 2012ASSL..384.....D, 2013A&A...550L...7G,2014ARA&A..52..487S,2016ApJ...817...66K, 2016JPhCS.728b2008D, 2018Natur.561..498J,2023A&A...678A..55R, 2024ApJ...976..125D,2024ApJ...974..270C,2025AJ....169..128S, 2025PASP..137j4201M}. The intense mass loss experienced by LBVs halts their progression toward cooler temperatures, preventing them from evolving into red supergiants (RSGs) \citep{1994PASP..106.1025H}. These LBVs are losing mass by high mass-loss rates, up to approximately $\rm 10^{-6}\, to \, 10^{-3}~\rm M_{\odot}~yr^{-1}$ \citep[e.g.,][]{1997ASPC..120...58L, 2015ASSL..412...77V, 2020Galax...8...20W}. During this phase, their bolometric luminosity increases for several months to years, accompanied by rapid mass loss at even higher rates \citep{2016ApJS..222....8D, 2016ApJ...817...66K}. The theoretical framework exists to describe this mass loss through super-Eddington continuum-driven winds \citep[e.g.,][]{2004ApJ...616..525O, 2006A&A...446.1039K,2014ARA&A..52..487S}.

During these mass loss events, LBVs can lose up to 10-20 $\rm M_{\odot}$ of material over a short time period, often accompanied by a dramatic increase in luminosity. The most famous example of a Giant Eruption (GE) is $\eta$ Carinae, which underwent a giant outburst in the mid-$\rm 19^{\rm th}$ century \citep[e.g.,][]{1997ARA&A..35....1D, 2007MNRAS.378.1609K, 2008NewA...13..569K, 2012Natur.486E...1D, 2012ASSL..384....1H, 2006MNRAS.372.1133G, 2016ApJ...817...66K, 2016ApJ...825..105K, 2017RSPTA.37560268S}. This GE ejected $\gtrsim \rm 10~M_{\odot}$ of material and created the Homunculus Nebula, a structure that remains visible till today. Several theories have been proposed to explain the mechanisms driving these outbursts within their stellar envelopes. One proposed physical explanation is given by 
\citet{2001MNRAS.325..584S}, who argued that the tidal interaction with a companion star could provide the energy to the envelope, making it unstable and triggering the GEs. 
Another explanation of these outbursts is opacity-driven super-Eddington luminosity in the stellar envelope \citep[e.g.,][]{2004ApJ...616..525O, 2016MNRAS.458.1214Q}, which causes the star to become unstable.
The impact of the GEs on massive stars remains a key question. \citet{2009NewA...14..539H} simulated a GE for a 190 $\rm M_{\odot}$ star, by using the mass loss rate of $\rm 1~M_{\odot}~yr^{-1}$ over a period of 20 years. Their finding suggested that rapid mass loss is triggered by short-timescale causes, leading to fluctuations in luminosity and radius. However, post-eruption modeling was investigated by \citet{2016ApJ...817...66K} and he used energy-conserving methods to study recovery, showing that mass ejection was driven by $\kappa -$ mechanism-induced radial pulsations. It has been proposed that eruptive LBV-like events are connected to the SNe IIn \citep[e.g.,][]{2008ApJ...679.1467S, 2007ATel.1263....1G}. These LBV-stars are among the most massive and luminous stars in the universe, characterized by their extreme instability. Despite experiencing such intense outbursts, most LBVs persist for centuries without undergoing a SNe explosion.

Observations of supernovae (SNe) IIn \citep{1994ApJ...420..268C,1997ARA&A..35..309F, 2010ApJ...709..856S, 2012AJ....144..131Z, 2019ApJ...872..135K, 2024ApJ...977..152F}, suggests that some massive stars undergo episodes of intense mass loss in the months to years leading up to their explosion. These mass-loss events are often characterized by the ejection of large amounts of material at high rates, which can reach up to $\rm 10^{-4}~M_{\odot}~yr^{-1}$ \citep[e.g.,][]{1988SSRv...46..225D, 1994ApJ...428L..17C, 1997ARA&A..35..309F, 2004MNRAS.352.1213C, 2007ApJ...659L..13O, 2010ApJ...724.1396O, 2013Natur.494...65O, 2009ApJ...695.1334S, 2012ApJ...744...10K, 2016MNRAS.456..853P, 2022ApJ...924...15J, 2024A&A...686A.231R, 2024ApJ...964..181H}. The expelled material forms a dense circumstellar medium (CSM) around the progenitor star. When the SNe shockwave interacts with this dense CSM, it produces the strong, narrow emission lines in the spectra that are characteristic of Type IIn SNe \citep[e.g.,][]{1994ApJ...428L..17C, 2012AJ....144..131Z, 2023MNRAS.519.1618Y, 2024MNRAS.530..405S}. To date, there are many SNe whose precursor outbursts were seen before the SN explosion. For e.g., Ibn SN2006jc \citep[e.g.,][]{2007ApJ...657L.105F, 2008MNRAS.389..113P} , the type IIn SN2009ip \citep[e.g.,][]{2012ATel.4412....1S, 2013ApJ...767....1P, 2013ApJ...763L..27P, 2013MNRAS.430.1801M}, , type IIn SN2010mc \citep{2013Natur.494...65O}, SN 2014av \citep{2016MNRAS.456..853P}, IIn-P SN 2020pvb \citep{2024A&A...686A..13E}, IIn Supernova 2021qqp \citep{2024ApJ...964..181H}, and SN 2023fyq \citep{2024ApJ...977..254D}.

\citet{2014ApJ...785...82S} suggested that these pre-SNe outbursts are caused by the hydrodynamic instabilities in evolved massive stars, observed both within the Galactic and in nearby galaxies of the Local Group. These evolved massive stars, known to undergo episodes of violent, episodic mass loss, are often classified as LBVs. 
These objects show precursor eruptions detected as transients a few years before the SN explosions. SN 2006jc was the first to exhibit a brief outburst two years prior, with a 2004 eruption reaching a peak luminosity similar to  $\eta$ Carinae. Its post-SN spectrum featured narrow helium emission lines, revealing dense circumstellar material (CSM) \citep[e.g.,][]{2007ApJ...657L.105F, 2008MNRAS.389..113P}.
SN 2009ip underwent multiple similar eruptions over three years followed by another eruption (or explosion) in 2012, characterized by a double-peaked light curve and broad emission lines \citep[e.g.,][]{2012ATel.4412....1S, 2013ApJ...767....1P, 2013ApJ...763L..27P, 2013MNRAS.430.1801M}. Similarly, SN 2010mc \citep{2013Natur.494...65O}, had a double peak light curve was nearly identical to SN 2009ip. Since then more similar objects have been observed, most notably SN 2016bdu \citep{2018MNRAS.474..197P}, SN 2019zrk \citep{2022A&A...666A..79F}, SN 2023ldh \citep{2025A&A...701A..32P}, and SN 2023vbg \citep{2024TNSAN..22....1P}.

In \citet{2024ApJ...974..124M}, we simulated a Giant Eruption (GE) in a 70 $\rm M_{\odot}$ star during the post-main sequence (post-MS) phase, imposing a high mass-loss rate for 20 years. The star initially expanded, increasing its luminosity ($L$), but as mass loss continued, $L$ dropped sharply while the effective temperature ($T_{\rm eff}$) rose slightly. The expelled material caused fluctuations in the stellar radius ($R$), though the interior remained stable. The gravitational energy lost during the eruption was comparable to that of $\eta$ Carinae's GE in the 19$^{\rm th}$ century \citep{2012NewA...17..616S, 2013MNRAS.429.2366S}. Our results suggest that a higher mass-loss rate could shorten the star’s recovery time and potentially trigger another eruption, possibly influenced by a binary companion’s gravitational effects. Eventually, the star evolved toward cooler temperatures, redistributing mass and regaining equilibrium.

In this paper, we investigate the effects of MGEs on massive stars (at Z=0.02 and 0.008) by simulating an artificial eruption in their outer layers, following the approach of \citet{2024ApJ...974..124M}. We analyze the changes in the stellar characteristics and the star's evolutionary track on the HR diagram during the MGEs phases. Additionally, we study the star's recovery phase and evaluate the duration of its recovery period after every GE phase. 
The dynamical properties of a single mass eruption event have already been studied by \citet{2024ApJ...974..124M}. However,  observations of LBVs indicate that the mass eruption phase often occurs
more than once \citep[e.g.,][]{2010MNRAS.408..181P, 2010ATel.2897....1D, 2023MNRAS.521.1941A}. A mass eruption event
can alter the density structure of the envelope, and therefore the subsequent mass eruption event can show completely different observational features. From this viewpoint, we simulated the evolution of the eruption event three times in a row. Additionally, we consider the effect of metallicity during the MGEs. 

In  \citet[e.g.,][]{mukhija2025accretionrecoverygianteruptions, 2026NewA..12202475M}, we study how giant eruptions in massive binary systems affect the companion star through episodic mass accretion. Using numerical simulations, we show that low accretion rates $\rm \lesssim 0.01 ~M_{\odot}~\rm yr^{-1}$ cause minor luminosity changes, while higher rates $\rm \gtrsim 0.01 ~M_{\odot}~\rm yr^{-1}$ lead to significant luminosity increase, envelope inflation, and cooling. The star later recovers, mixes the accreted material, and evolves as a more massive object.

The paper is organized as follows. The basic assumptions and method of modeling the $\rm  100~M_{\odot} $ star are discussed in section \ref{2}.  The simulations and results are described in section \ref{3}. Our discussion and summary are given in section \ref{4} and \ref{5} respectively.

\section{MESA modeling}
\label{2}

This section provides an overview of the stellar parameters, and the physical mechanics to construct our models. The models presented in this study are produced by using the 1D stellar evolution code \textsc{mesa} (Modules for Experiments in Stellar Astrophysics; version r23.05.1), which has been referenced in various papers by Paxton \citep[e.g.,][]{2011ApJS..192....3P, 2013ApJS..208....4P, 2015ApJS..220...15P, 2018ApJS..234...34P, 2019ApJS..243...10P}.  

We model a non-rotating $\rm 100~M_{\odot} $ star with a metallicity of $ Z=0.02$ and $0.008$, and apply episodic mass loss to the outer envelope of the star over three cycles for a period of 1 year. After each cycle, we allow the star to recover and reach a new equilibrium state.  All models start with the following chemical composition, uniformly distributed throughout the star. The metal fraction, $Z$, and the mass fraction of individual metals scale according to the solar abundances from  \citet{1998MNRAS.298..525P}. The helium mass fraction $Y$,  is calculated using $Y = Y_{\rm prim} + (\Delta  Y / \Delta  Z) / Z $  for a given $Z$. The values we used in our model are $ Y_{\rm prim}=0.24$ and $(\Delta  Y / \Delta  Z) / Z =2$. Both follow the default values in \textsc{mesa} \citep{10.1046/j.1365-8711.1998.01658.x}. The hydrogen mass fraction is given by $X = 1-Y-Z$.

The core region of massive stars is unstable to convection due to the enormous amount of energy generated in the center. In our model, we employ the standard Mixing Length Theory (MLT) \citep{1958ZA.....46..108B,1965ApJ...142..841H} with a fixed mixing length parameter of 1.6.  The same approach is used for the subsurface convection layers of the star. It determines the convective boundary using the Ledoux criterion \citep[e.g.,][]{1952ApJ...116..317O, 1975ApJ...195..157C}. In our calculations, the MLT++ approach is disabled, meaning that the temperature gradient predicted by MLT++ is used as is, without reducing it in the superadiabatic layers. Generally, it uses to counteract the inflation during the post-MS phase\citep[e.g.,][]{1999PASJ...51..417I, 2006A&A...450..219P, 2012A&A...538A..40G, 2015A&A...580A..20S, 2015ApJ...813...74J}, and leads the star's evolution towards the hotter side of the HR diagram \citep[e.g.,][]{2011ApJS..192....3P, 2021MNRAS.506.4473S, 2022MNRAS.514.3736S, 2022A&A...668A..90A}. \textsc{mesa} applies a diffusive treatment for semiconvection, with the diffusion coefficient based on \citet{1983A&A...126..207L} approach. We set a fixed semiconvection factor of 1 \citep[e.g.,][]{2006A&A...460..199Y, 2019A&A...625A.132S}. Additionally, convective overshooting is included with the overshooting parameter, \textbf{$\rm f_{ov}=0.01$ } \citep{2000A&A...360..952H}. We follow the default prescription in MESA, exponential overshooting above the core, and did not explicitly select any specific burning region. In our model, we employ the atmospheric (\texttt{atm}) module to compute the surface temperature and pressure, reflecting the conditions at the base of the stellar atmosphere. As the star's interior evolves, these values are used as boundary conditions for the simulation. We set the \texttt{atm} option to employ the $T-\tau$ formalism and choose the Eddington grey approximation to define the temperature distribution in the atmosphere. As we employ an artificial mass loss mechanism to induce MGEs in the outer envelope of the star, we use a time step shorter than the dynamical timescale of the star to capture the envelope's dynamical response. We employ the \texttt{Dutch} prescription to model the hot, line-driven stellar winds during the recovery phases only. This prescription combines several wind mass-loss models based on the $T_{\rm eff}$ and surface hydrogen abundance \( X_{\rm H}\). For models with \( T_{\rm eff} > 10^4~\mathrm{K} \) and  \( X_{\rm H} > 0.4 \), it use the prescription of \citet{2001A&A...369..574V}. For \( T_{\rm eff} > 10^4~\mathrm{K} \) and \( X_{\rm H} < 0.4 \), corresponding to H-poor stars it adopt the formulation of \citet{2000A&A...360..227N}. We evolve a non-rotating star to isolate the effects of mass loss enhanced by rotation on the star's evolution \citep[e.g.,][]{1998A&A...329..551L, 2000ARA&A..38..143M, 2012ARA&A..50..107L, 2014A&A...564A..57M}.
For the reaction network, we use the basic.net framework, which includes eight isotopes. \textsc{mesa} utilizes the default OPAL type I opacity tables from \citet{1993ApJ...412..752I, 1996ApJ...464..943I}, which assume fixed metal distributions. However, significant variations in the metal fractions during stellar evolution can alter the opacity of the envelope. Thus, we adopt the OPAL type II opacity tables \citep{1996ApJ...464..943I} to account for time-dependent changes in metal abundances in our model.

\begin{table*}
    \begin{tabular}{l | c c  c c c c c}
\hline
\hline
{ $Z$ = 0.02 } & & \multicolumn{4}{c}{Model 1: $\rm 10^{-2}~M_{\odot}~yr^{-1}$} \\
  \hline  
Stellar parameter & Point B & Point C & Point B' & Point C' & Point B''& Point C'' & Point B''' \\
\hline
\hline
 $\rm star~age~$($\rm M yrs$) &  2.380614 & 2.380615 & 2.380619 & 2.380620 & 2.380633 & 2.380634 &  2.380641      \\
 $\Delta t ~($\rm  yrs$)$ & -& 1 & 3.71 & 1 & 3.25 & 1 & 7.19\\

      $ M ~( \rm M_{\odot})$ & 100 & 99.99 & 99.989 & 99.979 & 99.979 & 99.969 & 99.962   \\
      $\log T_{\rm eff} ~(\rm K)$ & 4.30 & 4.40 & 4.30 & 4.40 & 4.30 &  4.40 & 4.30  \\
     $\log L~ (\rm L_{\odot})$ &  6.29 & 6.20 & 6.29 &6.21 &  6.29 & 6.21 & 6.29  \\
     $\log R ~ (\rm R_{\odot})$  & 2.07 & 1.83 & 2.07 & 1.83 & 2.07 &1.83 & 2.07   \\
     $\log  g ~ (\rm cm~s^{-2})$ & 2.29 & 2.79 & 2.30 & 2.79 & 2.30 & 2.79 &  2.30\\ 
     $\log \dot{M}~(\rm M_{\odot}~\rm yr^{-1})$      & - & -2.0 & -4.52 & -2.0 & -4.52 &  -2.0 & -4.52\\
     \hline 
     \hline
     \end{tabular}

\begin{tabular}{l | c c  c c c c c}
{ $Z$ = 0.02} & & \multicolumn{4}{c}{Model 2: $\rm 10^{-1}~M_{\odot}~yr^{-1}$} \\
  \hline  
Stellar parameter & Point P & Point Q & Point P' & Point Q' & Point P''& Point Q'' & Point P''' \\
\hline
\hline

     $\rm star~age~$($\rm M yrs$) &  2.380614 & 2.380615 & 2.380675 &2.380676 & 2.380725 & 2.380726 &  2.380767 \\
     $\Delta t ~($\rm  yrs$)$ & -& 1 & 40.9 & 1 & 49.04 & 1 & 59.91\\
     
     $ M ~( \rm M_{\odot})$ & 100 & 99.90 & 99.89 & 99.79 & 99.79 & 99.69 & 99.69 \\

     $\log T_{\rm eff} ~(\rm K)$ & 4.30 &4.44 & 4.30 & 4.44 & 4.30 &  4.44 & 4.30\\
     $\log L~ (\rm L_{\odot})$ &  6.29 & 5.99 &  6.29 & 5.99 & 6.29 &5.98 & 6.29  \\
     $\log R ~ (\rm R_{\odot})$  & 2.07 &1.63 & 2.07 &1.68 &   2.07 & 1.63 & 2.07 \\
     $\log  g ~ (\rm cm~s^{-2})$ & 2.29 & 3.18 & 2.30 &3.17 & 2.30 & 3.17 & 2.30  \\ 
     $\log \dot{M}~(\rm M_{\odot}~\rm yr^{-1})$      & - &-1.0 & -4.53 & -1.0 & -4.53 &-1.0 & -4.53 \\
     \hline 
     \hline
     \end{tabular}

    \begin{tabular}{l | c c  c c c c c}
{$Z$ = 0.008 } & & \multicolumn{4}{c}{Model 1: $\rm 10^{-2}~M_{\odot}~yr^{-1}$} \\
  \hline  
Stellar parameter & Point B & Point C & Point B' & Point C' & Point B''& Point C'' & Point B''' \\
\hline
\hline

 $\rm star~age~$($\rm M yrs$) & 2.569437 & 2.569438 & 2.569441 & 2.569442 & 2.569444 & 2.569445 & 2.569448 \\
 $\Delta t ~($\rm  yrs$)$ & -& 1 & 2.39 & 1 & 2.60 & 1 & 2.73\\
      $ M ~( \rm M_{\odot})$ & 100 & 99.99 & 99.98 & 99.97 & 99.97 & 99.96 & 99.96   \\
      $\log T_{\rm eff} ~(\rm K)$ & 4.30 & 4.34 & 4.30 & 4.34 & 4.30 &  4.34 & 4.30 \\
     $\log L~ (\rm L_{\odot})$ &  6.29 & 6.24 & 6.29 &6.24 &  6.29 & 6.24 & 6.29  \\
     $\log R ~ (\rm R_{\odot})$  & 2.07 & 1.95 & 2.07 & 1.95 & 2.06 &1.95 & 2.07    \\
     $\log  g ~ (\rm cm~s^{-2})$ & 2.29 & 2.52 & 2.30 & 2.53 & 2.31 & 2.53 & 2.30  \\ 
     $\log \dot{M}~(\rm M_{\odot}~\rm yr^{-1})$      & - & -2.0 & -4.79 & -2.0 & -4.78 &  -2.0 & -4.79 \\
     \hline \hline
     \end{tabular}
   
    \begin{tabular}{l | c c  c c c c c}
{$Z$ = 0.008} & & \multicolumn{4}{c}{Model 2: $\rm 10^{-1}~M_{\odot}~yr^{-1}$} \\
  \hline  
Stellar parameter & Point P & Point Q & Point P' & Point Q' & Point P''& Point Q'' & Point P''' \\
\hline
\hline

 $\rm star~age~$($\rm M yrs$) &  2.569437 &2.569438  & 2.569464 & 2.569465 &  2.569509 &    2.569510 & 2.569546 \\
 $\Delta t ~($\rm  yrs$)$ & -& 1 & 26.21 & 1 & 43.80 & 1 & 35.31\\
      $ M ~( \rm M_{\odot})$ & 100 & 99.90 & 99.89  & 99.79 & 99.79 & 99.69 & 99.69 \\
      $\log T_{\rm eff} ~(\rm K)$ & 4.30 & 4.38 &  4.30 & 4.38 & 4.30 & 4.38 &4.30\\
     $\log L~ (\rm L_{\odot})$ &  6.29 & 6.05 & 6.29 & 6.04 & 6.28 & 6.04 & 6.28 \\
     $\log R ~ (\rm R_{\odot})$  & 2.07 &  1.79 & 2.06 & 1.78 & 2.07 & 1.79 & 2.06 \\
     $\log  g ~ (\rm cm~s^{-2})$ & 2.29 & 2.86 &2.30 & 2.87 & 2.31 & 2.86 & 2.30\\ 
     $\log \dot{M}~(\rm M_{\odot}~\rm yr^{-1})$      & - & -1.0 & -4.79 & -1.0 & -4.79 &  -1.0 & -4.79\\
     \hline \hline
     \end{tabular}
    \caption{Stellar parameters shown above correspond to model 1, and model 2 for the Galactic and LMC metallicities at the end profiles of the eruption and recovery phases. Here the rows are: star age, period difference of eruption and recovery phase ($\Delta t$),  the mass of the star ($ M$), effective temperature ($ T_{\rm eff}$), surface luminosity ($ L$), surface radius ($ R $), surface gravity ($ g $), and mass loss rate ($ \dot{M}$) respectively.}
  
    \label{table1}
\end{table*}

\section{Results}
\label{3}

In this section, we systematically examine the effect of GEs on a \( 100~\mathrm{M}_\odot \) post-MS star for two different mass-loss rates, each undergoing three cycles of GEs at varying metallicities. Before initiating the eruptions, the models go through a pre-eruption evolutionary phase. For Model 1, as shown in Figure~\ref{HR}, panel~(a), this phase extends from the zero-age main sequence (ZAMS) (point A) to the post-MS phase (point B). For Model 2, shown in Figure~\ref{HR}, panel~(b), the evolution proceeds from the ZAMS (point A) to the post-MS phase (point P). During these pre-eruption phases in both models, we do not employ any wind mass-loss prescription. As a result, the stellar mass at points B and P remains \( 100~\mathrm{M}_\odot \). In addition to the eruption phases, we also analyze the recovery phases in detail.

In this section, we systematically examine the effect of GEs on the $\rm 100~M_{\odot}$ post-MS star for two mass loss rates, which undergoes three cycles of GEs at different metallicity values. Additionally, we also analyze their recovery phase in detail.

\subsection{Model setup:}
\label{3.1}

We evolve $\rm 100~M_{\odot}$ star, and introduce two artificial mass loss rates: $\rm 10^{-2}$ (model 1), and $\rm 10^{-1}~M_{\odot}~yr^{-1}$ (model 2), for 1 year during the three cycles of the GE. The GEs are initiated at the $T \simeq 19\,400$ K, followed by \citet{2024ApJ...974..124M}. After the first GE, we allow the star to recover, and it reaches a new equilibrium state during the recovery phase. Once the star returns to the point close to the point where the first GE is initiated, we trigger another GE for 1 year, followed by another recovery phase. Thus, this process is repeated for the three cycles of the GE, which we refer to as ``Multiple Giant Eruptions'' (MGEs). 
In Model 1, we induce a mass loss rate of $\rm 10^{-2}~M_{\odot}~yr^{-1}$ for 1 year during each of the three GE cycles. This results in a total mass loss of $\rm 0.06~M_{\odot}$ over the three cycles. Similarly, in Model 2, each eruption cycle results in a mass loss of $\rm 0.1~M_{\odot}$, with a total mass loss of $\rm 0.3~M_{\odot}$ after three cycles. Giant eruptions can eject more mass than that, but we stay at these conservative values as the results of MESA are more reliable and easier to converge for these mass loss rates. During each recovery phase, we apply the `Dutch' stellar wind mass loss prescription, which also leads to the mass loss due to the hot winds.

\subsection{Outcomes: Model 1 at Galactic metallicity}

\begin{figure*}
  \centering
  \begin{tabular}{c @{\qquad} c }
    \includegraphics[trim={0.5cm 0.0cm -0.3cm 0.2cm},clip,width=0.5\textwidth]{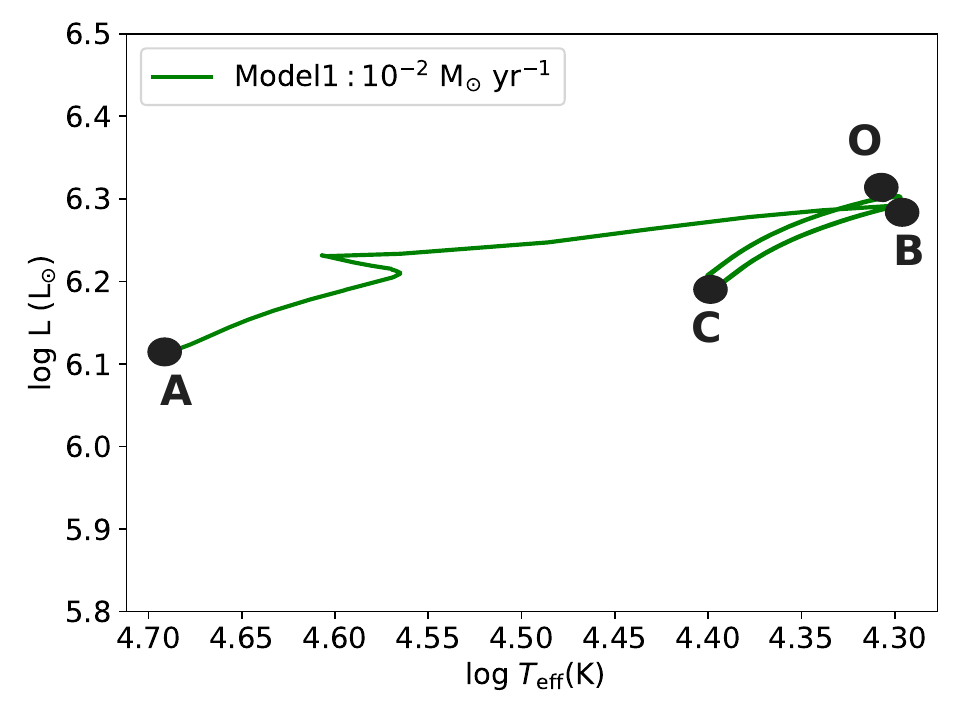}
    
    &
     \hspace{-0.9cm} 
    \includegraphics[trim={0.2cm 0.0cm 0.0cm 0.2cm},clip,width=0.5\textwidth]{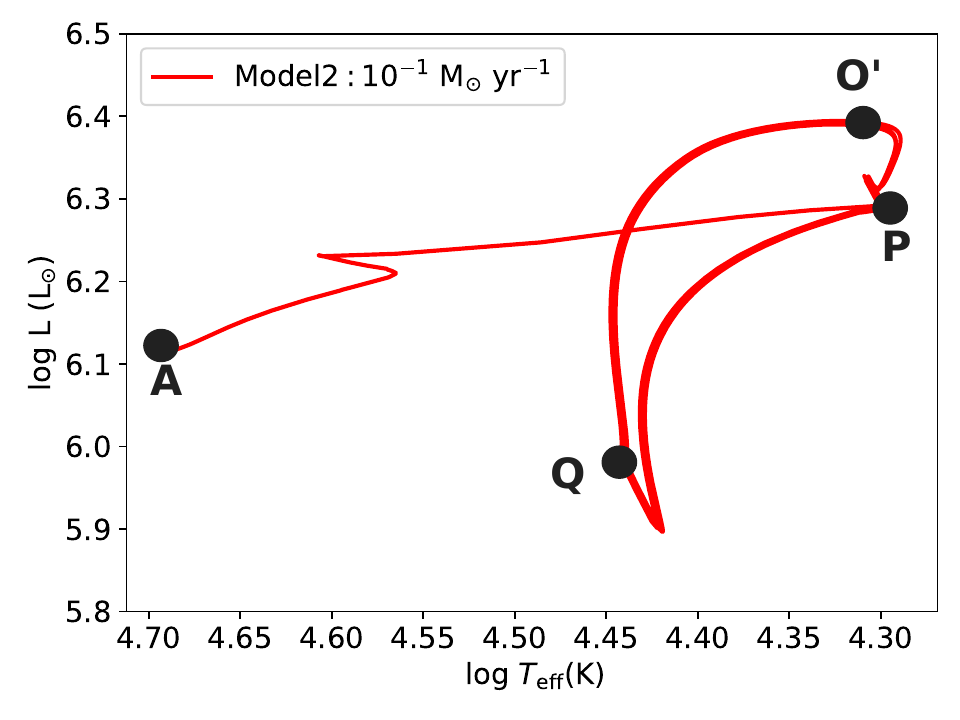}
    \\
    \small (a) & (b)
  \end{tabular}

  \caption{The stellar tracks of luminosity and temperature (HR diagram) for a $\rm 100~M_{\odot}$ star with Galactic metallicity, obtained using \textsc{mesa}, are shown. Here, A represents the zero-age-MS phase in both panels. Panel (a) depicts the evolution of a $\rm 100~M_{\odot}$ star undergoing three cycles of giant eruptions with a mass loss rate of $\rm 10^{-2}~M_{\odot}~yr^{-1}$ over 1 year, including the recovery phases (Model 1). Panel (b) shows the evolution of a $\rm 100~M_{\odot}$ star undergoing three cycles of giant eruptions with a mass loss rate of $\rm 10^{-1}~M_{\odot}~yr^{-1}$  over a 1 year period, along with the recovery phases (Model 2). Here, the evolutionary tracks during the three cycles of GE in both panels are overlapping, so we only show the first GE for both models. Additionally, in both models, the first GE is initiated at the same location but is labelled differently: in model 1, it corresponds to point B, while in model 2, it corresponds to point P.}
  \label{HR}
  \end{figure*}

In Model 1, at point B (Figure \ref{HR} panel (a)), we introduce a mass loss rate of $\rm 10^{-2}~M_{\odot}~yr^{-1}$ for a period of 1 year (representing the first GE). The evolutionary track from point B to C in Figure \ref{HR} panel (a) shows the result of this first GE. As a result of this mass loss, the star evolves towards the hotter side of the HR diagram, shedding its outer, hotter, and deeper layers. Initially, the luminosity increases for a period of $\rm \approx 8.5$ days (track from point B to O), and it surpasses the Eddington luminosity. Later on, it is followed by a sharp drop over the rest of the year, as shown in the evolutionary track from O to C. The luminosity variation, characterized by an initial increase followed by a sharp decline, aligns with the behavior described by \citet{2024ApJ...974..124M}. However, the duration of the initial luminosity rise in our model is more than two orders of magnitude longer than in \citet{2024ApJ...974..124M}, primarily due to differences in the mass loss rate and mass of the star. This suggests that a lower mass loss rate results in a smaller luminosity increase occurring over a longer timescale. Here,  the characteristic of luminosity depends only on the mass loss rate, not on the mass loss duration. At point C, the star becomes hotter and more contracted compared to point B. The stellar parameters corresponding to points B and C are given in Table \ref{table1}. The total gravitational energy lost during the first GE is $\rm 9 \times 10^{46}$ erg.
After this, the star undergoes a recovery phase, and it starts to evolve towards the cooler side of the HR diagram with almost increasing luminosity as mentioned in \citet{2024ApJ...974..124M} as well. Throughout this stage, the star's outer layers have adjusted, and it settles into a new quasi-equilibrium. Quantitatively, recovery is identified by comparing key stellar parameters at this phase: the luminosity remains slightly below the pre-eruption value by the order of 0.0024 dex, the radius is somewhat increased, and the effective temperature is nearly unchanged. This combination of structural stability and minimal deviation in physical properties signals the end of the recovery phase. Thus, it returns to a state which is very near to the initial point where the first GE is initiated. The track from the point C to B' illustrates the first recovery phase as shown in Figure \ref{LTR} (we refer to the Figure \ref{LTR} as we did not mark these points on the Figure \ref{HR} due to the overlapping of the evolutionary tracks). Interestingly, during this recovery, the star stabilizes at $\log L$, a value significantly lower than its Eddington luminosity ($L_{\rm Edd}$) at the end phase of the recovery. The stellar parameters corresponding to point B' are provided in Table \ref{table1}. It took 3.71 years for the star to evolve from point C to B', which is referred to as the first recovery period of the star.

At point B', we initiate the second GE, and the evolutionary track from point B' to C' in Figure \ref{LTR} illustrates this phase. A mass loss rate of $\rm 10^{-2}~M_{\odot}~yr^{-1}$ is applied over 1 year, similar to the first GE. The position, where we initiate the second GE, is consistent with that of the first GE ($T = 19\,400$ K). As a result of the mass loss, the evolutionary track shifts towards the hotter side, accompanied by a drop in luminosity, similar to the first GE. During this time, initially luminosity increases for a period of $\approx 7.6 $ days. The total gravitational energy lost during this 1 year is $\rm 9 \times 10^{46}$ erg.
At point C', we let the star evolve further, and it reaches a new equilibrium state, referred to as the second recovery phase. The evolutionary track from C' to B'' in Figure \ref{LTR} represents this phase.  The star takes 13.25 years to return to the point, very close to the point where the  GE is initiated. This is referred to the second recovery period of the star.  At point B'', the star stabilises with a lower luminosity, similar to the first recovery phase, while maintaining a consistent temperature. The stellar parameters corresponding to the points C', and B'' are listed in Table \ref{table1}.

 At point B, we initiate the third GE over 1 year. The evolutionary track from B'' to C'' in Figure \ref{LTR} illustrates this phase, which exhibits behavior similar to the first and second GEs. During this 3rd GE, initially luminosity increases for a period of $\approx 8.03$ days. The total gravitational energy lost during this eruption is $\rm 9 \times 10^{46}$ erg, and the stellar parameters at this stage are provided in Table \ref{table1}. Finally, at point C'', the star undergoes the third recovery phase. The evolutionary track from C'' to B''' in Figure \ref{LTR} represents this phase, with the recovery taking 7.19 years. The stellar parameters for this final recovery phase are detailed in Table \ref{table1}.

It is intriguing to observe that during the GEs, the initial increase in luminosity and its duration do not follow a specific pattern. Similarly, during the recovery phase after each GE, the time required for the star to reach the point where GEs are initiated varies, with the second GE having the longest recovery period and the first GE the shortest. While the loss of gravitational energy remains consistent since the same amount of mass is removed in each GE. This suggests that the evolving envelope structure after each eruption introduces variability in the early expansion and recovery period.\\

\subsection{Outcomes: Model 2 at Galactic metallicity}

\begin{figure*}
\includegraphics[trim= 0.0cm 0.0cm 0cm 0.0cm,clip=true, width=1\textwidth]{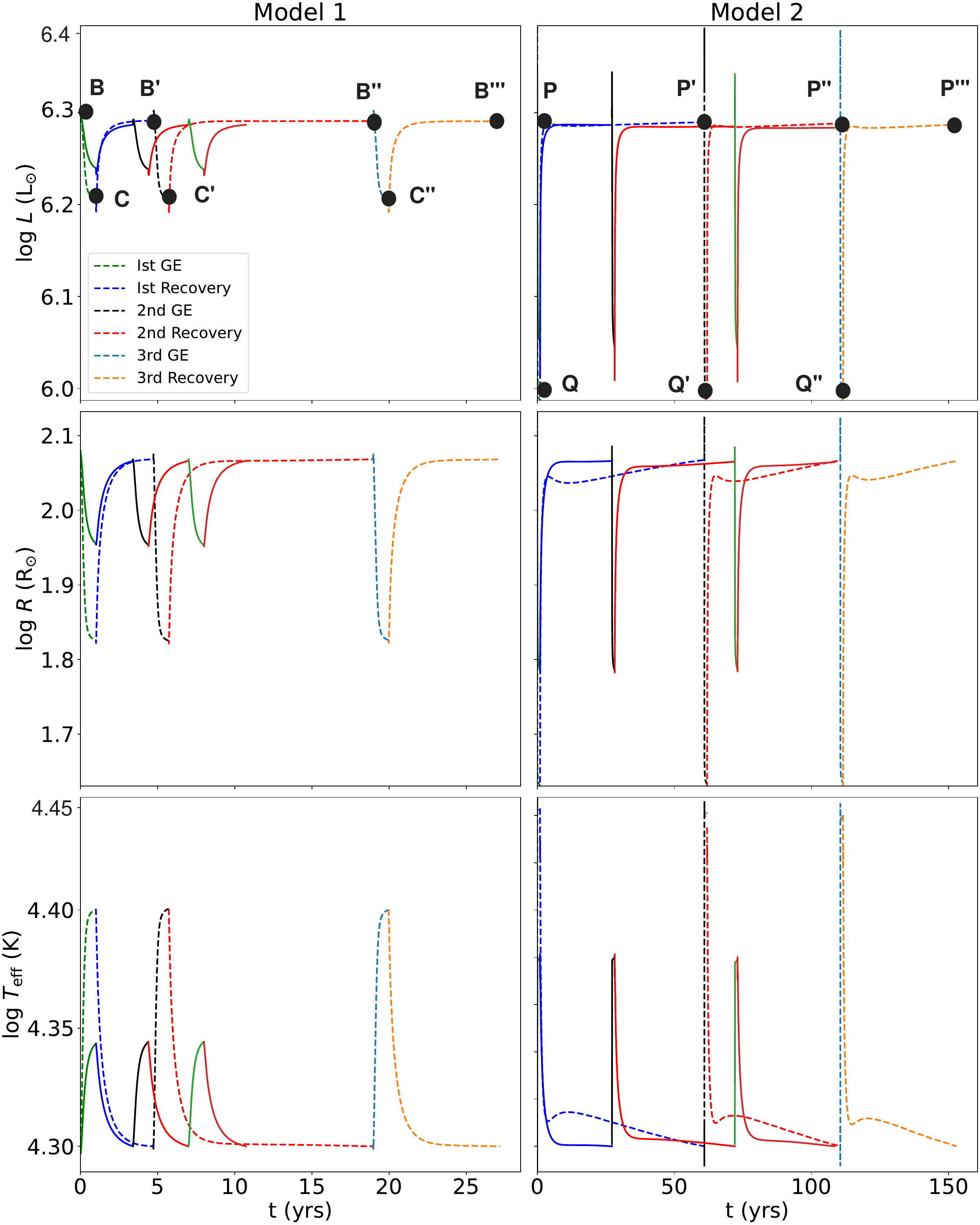} 
  \centering
\caption{The variation in luminosity $L$ (panel a), radius $R$, effective temperature $T_{\rm eff}$ (panel c) of the star during the MGEs, and recovery phases are shown for both model above. Here, the dashed line represents the GEs and recovery phases for $Z$ = 0.02, while the solid line represents the GEs and recovery for at $Z$ = 0.008. }
\label{LTR}
\end{figure*}

In Model 2, at point P (see Figure \ref{HR} panel (b), equivalent to point B, but labelled differently to avoid confusion), we introduce a mass loss rate of $\rm 10^{-1}~M_{\odot}~yr^{-1}$ over 1 year. The evolutionary track from point P to Q in Figure \ref{HR} panel (b) represents the first GE of Model 2. As a result of the induced mass loss, the star evolves toward the hotter side of the HR diagram, expelling hotter and deeper layers similar to Model 1. This behavior is consistent with Model 1. Initially, the luminosity increases over 5.47 days (track from point P to O'), followed by a sharp decline for the rest of the year. A comparison between Model 1, Model 2, and \citet{2024ApJ...974..124M} suggests that the duration of the early luminosity increase depends not only on the mass loss rate but also on the total mass of the star. The total gravitational energy lost during the first GE is $\rm 9 \times 10^{47}$ erg. The loss of gravitational energy is 10 orders of magnitude higher than in Model 1, which is expected since the ejected mass is also 10 orders of magnitude higher compared to Model 1. Afterwards, the star evolves to the recovery phase. The evolutionary track from point Q to P' in Figure \ref{LTR} represents the first recovery phase, which lasts for 59.91 years. Table \ref{table1} lists the stellar parameters corresponding to the points P, Q, and P'.

At point P', we initiate the second GE. During this phase, the luminosity again surpasses the Eddington luminosity for 5.43 days before sharply dropping. The evolutionary track from P' to Q' in Figure \ref{LTR} shows this second GE phase, with a total gravitational energy loss of $\rm 9 \times 10^{47}$ erg. From the point Q', the star evolved through a second recovery phase, represented by the evolutionary track from Q' to P'' in Figure \ref{LTR}. The recovery period during this phase is 49.04 years. The stellar parameters corresponding to the points Q', and P'' are also given in Table \ref{table1}.

At point P'', the third GE is triggered, over 1 year. The evolutionary track from P'' to Q'' in Figure \ref{LTR} shows this phase, with a total gravitational energy loss of $\rm 9 \times 10^{47}$ erg. During this phase, the initial luminosity surpasses the Eddington luminosity for 5.49 days. Following the third GE, the star undergoes its third recovery phase, with the evolutionary track from Q'' to P''' in Figure \ref{LTR} representing this phase. The total duration of this recovery is 40.9 years, which is much lower compared to the first and second recovery periods.  Later on, during this phase, the star expands and cools, similar to other recovery phases. In this model as well, the duration of the initial luminosity increase and the recovery period does not follow a specific pattern after comparing with other models (see below Sect.). However, the duration of the initial luminosity rise remains nearly constant for each eruption, while the recovery period is the highest for the first GE after that, and it gradually decreases in subsequent recovery phases. This highlights the crucial role of the mass loss rate in shaping the star’s structure during both the eruption and recovery phases. The stellar parameters corresponding to the points Q'', and P''' are given in Table \ref{table1}.

\begin{figure}
    \centering
    \includegraphics[width=1\linewidth]{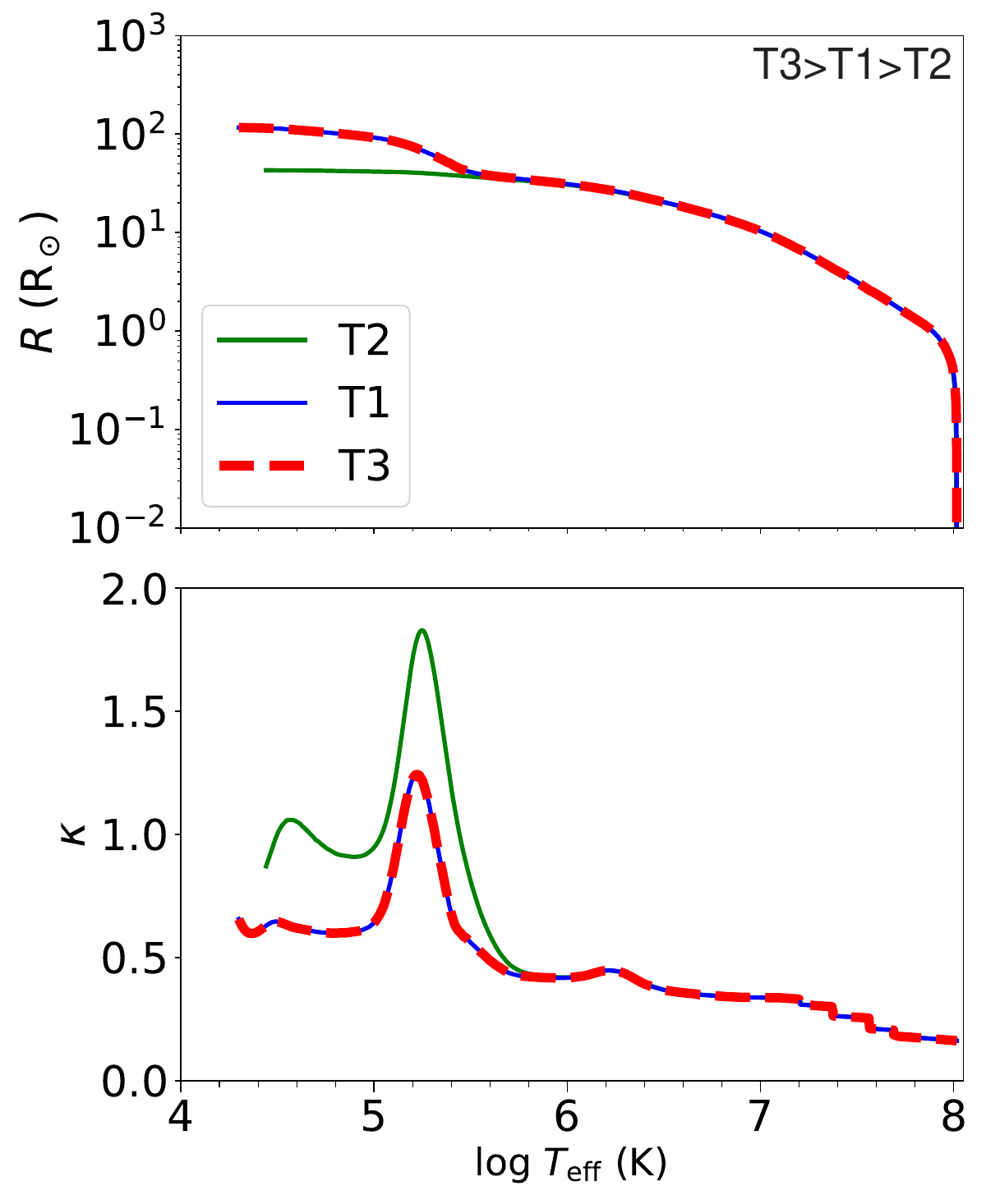}
    \caption{Figure shows the variation in the temperature from the outer layers to the interior of the star. During the recovery phase, contraction occurs, and thus the temperature and radius decrease.  Although the temperature changes are minimal, they lead to a significant increase in opacity, as shown in the bottom panel, in the outer layers at point T2 due to the thermal imbalance.}
    \label{opal}
\end{figure}
 
Additionally, we also find that during the recovery
phases of Model 2, just before reaching the temperature 
$19 \, 4000$, the star undergoes a contraction phase. It becomes hotter, and its luminosity slightly drops before
resuming further expansion. While this effect is not dominant, the contraction is noticeable. During the recovery phases of Model~2 at Galactic metallicity, just before reaching a temperature of \(19\,400~\mathrm{K}\), the star undergoes a brief contraction, as shown in the upper panel of Figure~\ref{opal}. The point T1 marks the onset of contraction in the recovery phase, occurring near \(\log T_{\rm eff}=4.31\). The point T2 represents the peak of contraction at  \(\log T_{\rm eff}=4.32\), while T3 corresponds to the final profile of the recovery phase at 
\(\log T_{\rm eff}=4.30\), which serves as our stopping condition, i.e., the end point of the third recovery phase. Consequently, the stellar radii at these stages follow  \(R_2 > R_3 > R_1\). As a result of contraction, the star becomes hotter and its luminosity experiences a slight decrease before continuing its expansion. The duration of this contraction phase is approximately \(5.5~\mathrm{yrs}\). We interpret this behavior as a consequence of \textit{thermal imbalance} in the stellar envelope during the recovery phase. Following the high mass-loss episode, the envelope departs from thermal equilibrium: radiative energy loss outpaces local heating, causing a temporary reduction in thermal pressure support and leading to a brief contraction. As the outer layers contract, density and temperature rise, increasing the local opacity (as seen in the lower panel of Fig.~\ref{opal}). Radiation becomes partially trapped, and as the star redistributes this energy, thermal pressure is gradually restored, allowing the envelope to resume its expansion. This 
thermally driven adjustment, seen only in Model~2 at Galactic metallicity and high mass-loss rates, suggests that strong winds can destabilize the envelope and induce oscillatory behavior during the recovery phase. Such thermal imbalance may help create favorable conditions for a subsequent eruption of the star.

\subsection{Outcomes: at LMC metallicity}

We also simulate these MGEs at a metallicity of $Z$ = 0.008 for both models. The evolutionary track behavior is similar to that at Galactic metallicity for both models.  However, the duration of the initial luminosity increase differs. For the Model 1, at lower metallicity, the star maintains luminosities exceeding the Eddington limit for a longer period, with durations of 12.95, 12.67, and 12.38 days for the first, second, and third GEs, respectively. The recovery phase at lower metallicity is also shorter compared to Galactic metallicity, lasting 2.39, 2.6, and 2.73 years for the first, second, and third recovery phases, respectively. The gravitational energy loss during each GE remains consistent at $3 \times 10^{46}$erg, which is a factor of $\sim 3$ lower than in Model 1 at Galactic metallicity. Similar to the Model 2 at the LMC metallicity, the star exhibits a comparable evolutionary track behavior. The initial luminosity increase lasts for a few days, as observed earlier. The duration of this period, for the first, second, and third GEs, is 8.5, 8.15, and 8.20 days, respectively, aligning with our previous findings that lower metallicity results in a prolonged period for the luminosity increase. Likewise, the recovery phases at lower metallicity last 26.21, 43.80, and 35.31 years, further supporting the consistency of our earlier results.

This suggests that for GEs, the initial rise in luminosity depends on the mass-loss rate, the mass of the star, and its composition. Similarly, the recovery period is also influenced by composition, with lower metallicity stars exhibiting a shorter recovery period. As a result, stars' eruptions occurring in lower metallicity environments tend to recover more quickly than those in higher metallicity environments. The stellar parameters correspond to these MGEs, and the recovery phase is mentioned in \ref{table1}. Additionally, Figure \ref{LTR} shows that the evolutionary track at Galactic and LMC metallicity suggests a smaller decrease in luminosity for the same mass loss rate and one year of duration in the LMC compared to the Galactic case. This indicates that the decline in luminosity also depends on the composition.

\subsection{Dependency of Luminosity Decline on Mass Loss Rate and Metallicity}

\begin{figure}
    \centering
    \includegraphics[trim= 0.0cm 0.0cm 1.0cm 1.0cm,clip=true,width=1.02\columnwidth]{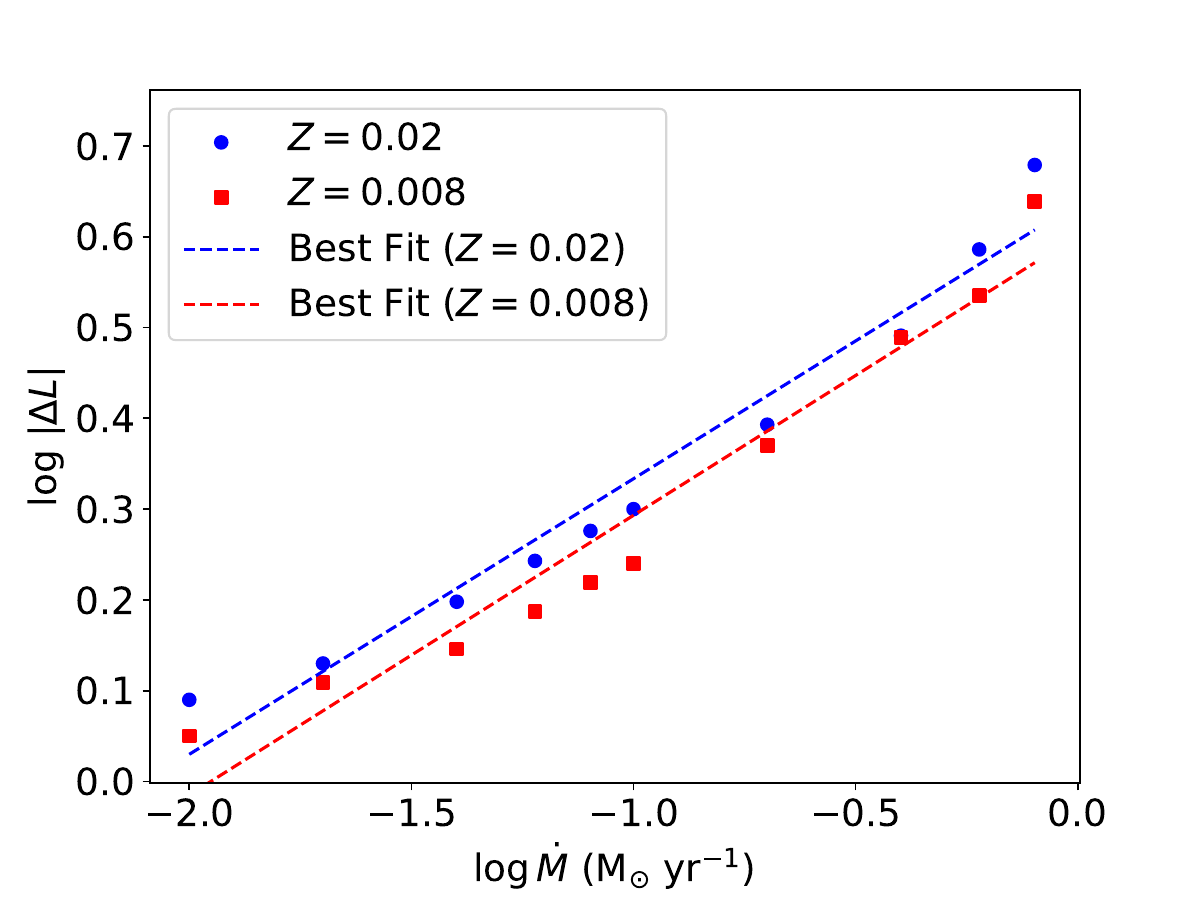}
    \caption{The relation between the mass loss rate during the eruption $\log \dot{M}$ and the decline in luminosity $\log |\Delta L|$ for two metallicities: $Z$ = 0.02 (blue) and $Z$ = 0.008 (red). The dashed lines represent the best-fit lines for each metallicity (see equation \ref{eq1}).}
    
    \label{fig3}
\end{figure}

We examine the correlation between mass loss rate  ($\dot M$) and the decline in luminosity during the eruption phase. As mass loss is introduced, we observe a corresponding decrease in luminosity along the evolutionary track as shown in Fig. \ref{HR}. To quantify this behavior, we simulated the $10$ mass loss rate ranging between $10^{-1}$ to 1 $\rm M_{\odot}~yr^{-1}$ over 1 year at the same location where we initiated earlier GEs. We examine this dependency at both metallicities: Galactic metallicity and LMC metallicity. %
 The results indicate that the magnitude of luminosity decline depends not only on the applied mass loss rate, as shown in Figure \ref{HR}, where Model 1 has less decrease in the luminosity value compared to Model 2. But it depends on the metallicity of the star as well as shown in Figure \ref{LTR}. At the higher metallicities, eruptions are more efficient, thus, the luminosity drop tends to be more pronounced compared to lower metallicity stars, where weaker eruptions result in a more gradual decline. We fit a power law relation between the variation in the luminosity due to the first GE over a period of 1 year by raging the mas loss rates between $10^{-2}$ to 1 $\rm M_{\odot}~yr^{-1}$. Thus the relation between the decline in luminosity and the mass loss rate for both metallicities is
\begin{equation}
\log |\Delta L|=
\begin{dcases}
\begin{aligned}
0.304^{ \pm 0.022} \log(\dot{M}) + 0.637 ^{\pm 0.025};\\
Z = 0.02
\end{aligned}\\
\begin{aligned}
0.325 ^{\pm 0.024}  \log(\dot{M}) + 0.625 ^{\pm 0.028};\\
Z = 0.008,
\end{aligned}
\end{dcases}
\label{eq1}
\end{equation}
where $\dot{M}$ is in units of $\rm M_{\odot}~yr^{-1}$. Fig. \ref{fig3} shows the extended set of 10 models, we ran to explore the luminosity response across a broader range of initial conditions and metallicities. The data points in Fig. \ref{fig3} align well with the power-law relations described in Equation \ref{eq1}. Notably, the dispersion around the best-fit line is slightly larger at lower metallicity, suggesting a  less efficient coupling between mass loss and luminosity change in such environments. This might be attributed to the reduced efficiency of stellar eruptions in low-metallicity stars, which can delay or moderate the envelope response to a given $\dot{M}$. It also highlights that, while the overall trend is consistent across models, there are modest deviations at the highest mass-loss rates, where the assumption of steady-state or eruptions structures may break down. In these cases, the mass ejection can drive shock-heated zones or trigger partial envelope detachment, leading to more abrupt luminosity changes than predicted by the power law alone. These features hint at the nonlinear hydrodynamic response of the outer layers under extreme conditions.

\subsection{Impact of MGEs on Stellar Surface Properties:}

\begin{figure*}
 
  \includegraphics[trim= 0.0cm 0.0cm 0cm 0cm,clip=true,width=0.9\textwidth]{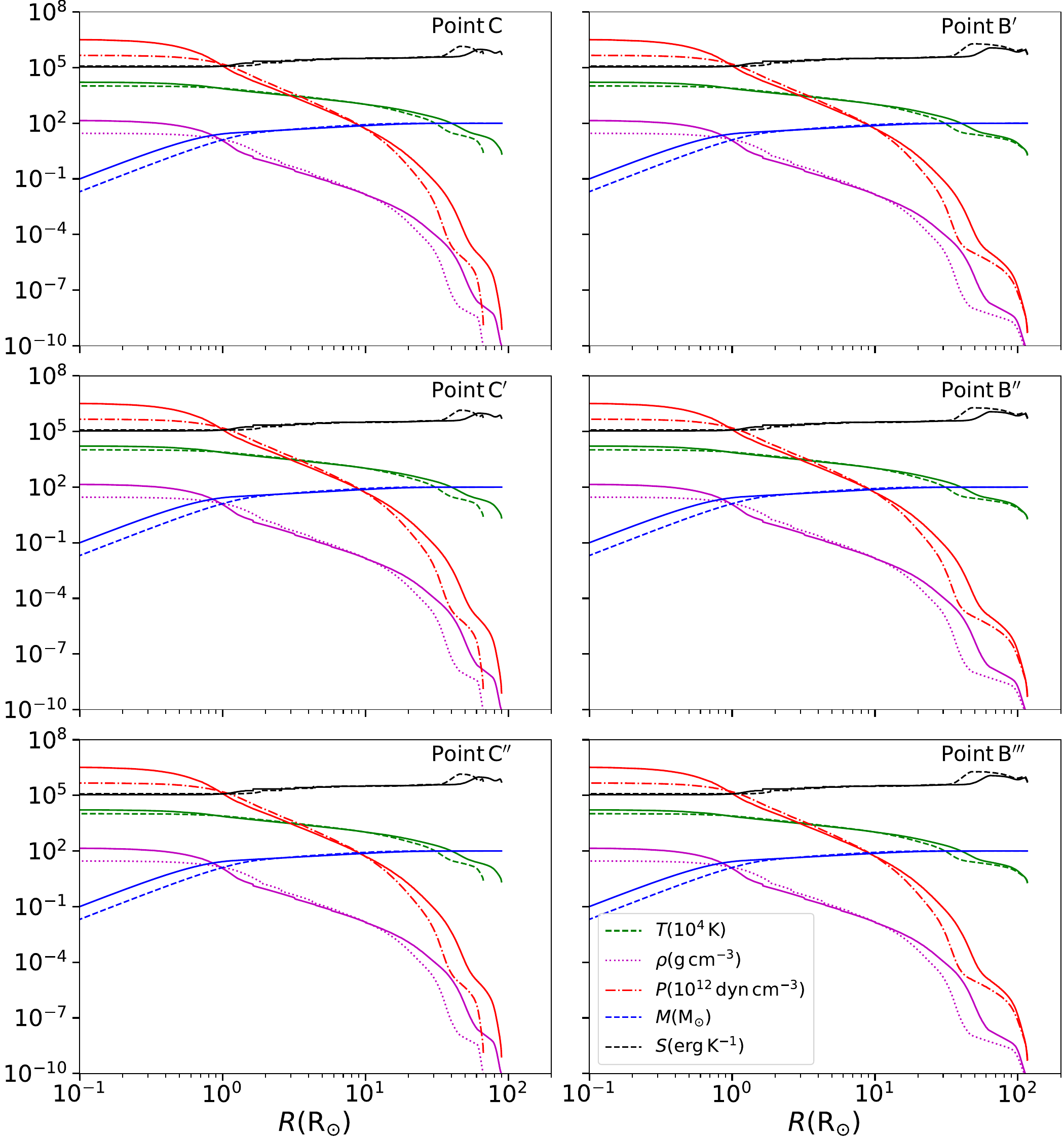} 
  \centering
  
\caption{The variation in the stellar properties, including temperature, density, pressure, mass, and entropy of a $ \rm 100~M_{\odot}$ during the MGEs, and recovery phase are shown above for Model 1. Here points, C, B', C', B'', C'', and B'''  represent the last profile of 1st GE, the last profile of 1st recovery, the last profile of 2nd GE, the last profile of 2nd recovery, the last profile of the 3rd GE, and last profile of the 3rd recovery phase. Here, the dashed line represents the stellar parameters at the Galactic metallicity, while the solid line represents the stellar parameters at LMC metallicity.}
\label{mod1}
\end{figure*}

\begin{figure*}
 
  \includegraphics[trim= 0.0cm 0.0cm 0cm 0cm,clip=true,width=0.9\textwidth]{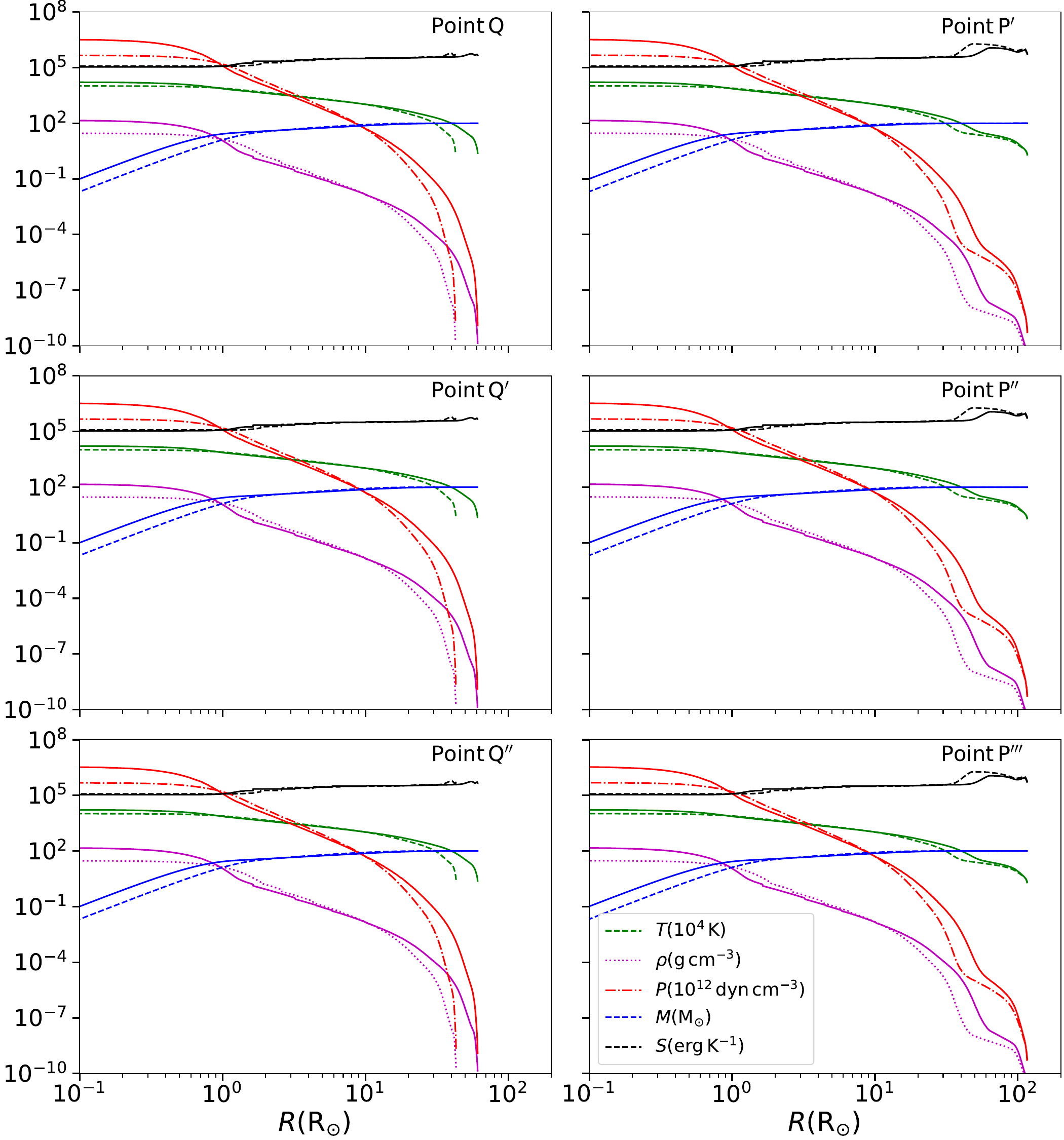} 
  \centering
  
\caption{The variation in the stellar properties, including temperature, density, pressure, mass, and entropy of a $ \rm 100~M_{\odot}$ during the MGEs, and recovery phase are shown above for Model 2. Here points, Q, P', Q', P'', Q'', and P'''  represent the last profile of 1st GE, the last profile of 1st recovery, the last profile of 2nd GE, the last profile of 2nd recovery, the last profile of the 3rd GE, and last profile of the 3rd recovery phase. Here, the dashed line represents the stellar parameters at the Galactic metallicity, while the solid line represents the stellar parameters at LMC metallicity.  }
\label{mod2}
\end{figure*}

Figure \ref{LTR} shows the variation in stellar parameters $L$, $R$, and $T_{\rm eff}$, during the GEs, and recovery phases at $Z$ = 0.02 and 0.008 for both Model 1 and Model 2, respectively. We observed that there are continuous drops in the luminosity track (row 1) during the eruption phases, and it increases during the recovery phases. The variations in the luminosity values for Model 2, during the eruption phase, are larger than those compared to Model 1, due to a higher mass loss rate. Additionally, the variation in the luminosity is lower at the LMC metallicity, compared to the galactic metallicity. This suggests that the metal-poor stars remain more structurally stable compared to metal-rich stars, as we removed the same amount of mass at both metallicity values. Row 2  depicts the variation in $R$, as the star contracts during the eruption, causing a decrease in $R$ value, and expands during the recovery phase, leading to an increase in $R$. Row 3 illustrates the temperature variation; the temperature rises during the eruption as the star expels its outer layers and shifts towards the hotter side, and decreases during recovery as the star evolves towards the cooler side of the HR diagram. Similar to the L, the star show less variation in R, $T_{\rm eff}$, at the lower metallicity.  

In Figures \ref{mod1}, and \ref{mod2}, the stellar properties including temperature ($T$), density $\rho$, pressure $P$, mass $M$, and entropy $S$ during the eruptions and recovery phases are shown from the outer envelope to the stellar interior of the star for Models 1 and 2 respectively at both metallicity. The inner regions of the star remain nearly constant, as the mass loss mechanism is implemented in the outer layers. During each eruption phase, the $T$, $\rho$, and $P$ in the outer layers are higher, compared to the recovery phase, as the star contracts during the eruption and expands during the recovery.

\section{Discussion}
\label{4}

We have carried out numerical simulations to
Investigate the properties of the repeated mass eruption events for a single $\rm 100~M_{\odot}$ star during the post-main-sequence phase. The key aspect of this work is that we induce mass eruption events not just once, as mentioned in \citet{2024ApJ...974..124M}, but three consecutive times by implementing an artificial mass-loss rate in the star's outer envelope. This approach aligns with the observed signatures of LBV stars and supernova pre-outbursts, where stars undergo multiple eruption phases rather than a single outburst event. In this work, we examine how a star's properties evolve following multiple eruption phases and their subsequent recovery periods. Additionally, we investigate the impact of metallicity on these eruption cycles. We used a publicly available stellar evolution code \textsc{mesa} to model the evolution of a $\rm 100~M_{\odot}$ star undergoing Multiple Giant Eruptions. We assume that this $\rm 100~M_{\odot}$ star is the primary in a binary system undergoing giant eruptions. However, we simulate it as a single star to isolate the effects of binary interactions. The simulation tracks the star through three cycles of eruptions, each eruption over a period of 1 year. After each eruption, we let the star evolve for recovery without any artificial mass loss. But we consider the hot wind mass loss during the recovery.  Once it approaches the point in its recovery evolution where the first GE is initiated, another eruption cycle is triggered. The decision to initiate these GEs at this specific stage i.e., $T_{\rm eff} = 19\,400 $ K is based on \citet{2010MNRAS.405.1924K}, that proposed that such systems are often binaries, where eruptions are induced by the gravitational influence of a companion in an eccentric orbit. According to this scenario, the timing of subsequent GEs could depend on the orbital position of the secondary, potentially delaying or even preventing future eruptions.

\section{Summary}
\label{5}
We examine two models for multiple giant eruptions in massive stars, Model 1 for the mass loss rate $10^{-2}$, and Model 2 for the mass loss rate $10^{-1}~\rm M_{\odot}~yr^{-1}$.
In both models, when a GE is initiated, the star undergoes a physical transition, evolving from the cooler to the hotter region of the HR diagram \citep{2012ApJ...751L..34V, 2022MNRAS.514.3736S, 2024ApJ...974..124M}. However, in Model 2, the luminosity decline is noticeably steeper compared to Model 1, as shown in Figure \ref{HR}. It is evident that higher mass loss leads to a more significant decrease in luminosity and a greater increase in temperature compared to Model 1. As a result of the GE, the star contracts, losing gravitational energy as material is expelled from its outer layers. This leads to a sharp drop in luminosity. During the eruption phase, in both models, the star exceeds the Eddington luminosity, with a rise in luminosity lasting several days, followed by a sharp decline for the rest of the year, and during the recovery phases, the star expands and cools as it evolves. The temperature evolution remains relatively constant, showing minimal variation in its values.
Thus for Model 1, during the first GE phase at Galactic metallicity, the luminosity decreases by a factor of 1.23, the radius decreases by a factor of 1.73, and the temperature increases by a factor of 1.25. Similarly, at LMC metallicity, the luminosity decreases by a factor of 1.12, the radius decreases by a factor of 1.31, and the temperature increases by a factor of 1.09. These values remain almost consistent for the second and third GE phases as well. For the Model 2, during the first GE phase at Galactic metallicity, the luminosity decreases by a factor of 1.99, the radius decreases by a factor of 2.75, and the temperature increases by a factor of 1.38. Similarly, at LMC metallicity, the luminosity decreases by a factor of 1.73, the radius decreases by a factor of 1.90, and the temperature increases by a factor of 1.20.

However, in Model 2, at Galactic metallicity, during all recovery phases, before reaching the final stage of recovery, the star undergoes contraction due to the thermal imbalance during the readjustments. This behavior is absent for Model 1, for both values of metallicity and for Model 2 at the LMC metallicity.

We evolve the star during the recovery phase until it reaches the point where the first GE is initiated, as shown in Figure \ref{LTR}. Later on, we also examine that if we evolve the star further beyond the end of the recovery phase, and it settles at a more luminous position on the HR diagram. This behavior is similar to the findings of \citet{2024ApJ...974..124M}, where a $\rm 70~M_{\odot}$ star settles at a higher luminosity after the recovery phase. However, in our case, the final position on the HR diagram is slightly more luminous, as the star in our models has a higher initial mass compared to the $\rm 70~M_{\odot}$ star studied by \citet{2024ApJ...974..124M}.

The simulated star behaves differently from observed GEs, which typically show a sustained luminosity increase. In contrast, our simulation exhibits an initial rise followed by a sharp decline in the luminosity, due to the energy driving the GE coming from accretion onto the companion star, rather than from the LBV itself \citep[e.g.,][]{2001MNRAS.325..584S, 2010ApJ...723..602K, 2016RAA....16..117S}. This is consistent with the accretion model for GEs where gravitational energy from accretion powers the GEs. Additionally, our model keeps the star's temperature high during the eruption, whereas observed GEs tend to be cooler \citep{1987ApJ...317..760D}. This is expected, as our model simulates only the star and excludes ejecta or a pseudo-photosphere.

We found that for Models 1 and 2, and radius decreases. Although the temperature changes each eruption cycle, the stellar surface exhibits similar behavior, with minimal variation in stellar parameters. However, significant differences arise during the recovery phase after each giant eruption (GE), particularly at higher mass loss rates. We also find out that during the GEs, the initial rise in luminosity and its duration do not follow a fixed pattern. Likewise, the recovery phase after each GE varies in duration for both models at both metallicity values. Although the loss of gravitational energy remains consistent since the same amount of mass is expelled in each GE. Thus, this suggests that the evolving structure of the stellar envelope after each eruption appears to introduce variability in both the early expansion and the recovery period. Additionally, we observed a consistent trend in lower-metallicity stars, where the recovery period remains more stable compared to those with higher metallicity. Notably, at lower metallicity, the outer layers do not contract during recovery, suggesting that stars with higher metallicity may be more susceptible to these eruptions due to opacity-driven mechanisms. The HR diagram for both models exhibits similar behavior across all MGEs.  

\vspace{0.5cm}
{We thank an anonymous referee for very helpful comments that helped to improve the paper.
We acknowledge the Ariel HPC Center at Ariel University 
for providing computing resources that have contributed to the research results reported in this paper.
BM acknowledges support form the AGASS center at Ariel University.
The MESA inlists and input files to reproduce our simulations and associated data products are available on Zenodo (DOI: 10.5281/zenodo.15286804).

\vspace{0.5cm}

\bibliography{GE_Massive}
\bibliographystyle{mnras}

\end{document}